\documentclass{article}

\usepackage{arxiv}

\usepackage[utf8]{inputenc} 
\usepackage[T1]{fontenc}    
\usepackage{url}            
\usepackage{booktabs}       
\usepackage{amsfonts}       
\usepackage{nicefrac}       
\usepackage{microtype}      
\usepackage{graphicx}
\usepackage{natbib}
\setcitestyle{numbers}

\usepackage{amsmath,amssymb}

\title{Optimal input representation in neural systems at the edge of chaos}

\author{
	Guillermo B. Morales\\
	Departamento de Electromagnetismo y F{\'i}sica de la Materia \\ Instituto Carlos I de F{\'i}sica Te{\'o}rica y Computacional.\\
	Universidad de Granada \\
	E-18071 Granada, Spain \\ 
    \texttt{guillermobm@onsager.ugr.es} \\
	\And
	Miguel A. Mu{\~n}oz \\
	Departamento de Electromagnetismo y F{\'i}sica de la Materia \\ Instituto Carlos I de F{\'i}sica Te{\'o}rica y Computacional.\\
	Universidad de Granada \\
	E-18071 Granada, Spain \\
	\texttt{mamunoz@onsager.ugr.es} \\
}

\date{}

\renewcommand{\shorttitle}{Optimal input representation in neural systems at the edge of chaos}

\begin{document}
\maketitle

\begin{abstract}
Shedding light onto how biological systems represent, process and store information in noisy environments is a key and challenging goal. A stimulating, though controversial, hypothesis poses that operating in dynamical regimes near the edge of a phase transition, i.e. at criticality or the "edge of chaos", can provide information-processing living systems with important  operational advantages, creating, e.g., an optimal trade-off between robustness and flexibility. Here, we elaborate on a recent theoretical result, which establishes that the spectrum of covariance matrices of neural networks representing complex inputs in a robust way needs to decay as a power-law of the rank, with an exponent close to unity, a result that has been indeed experimentally verified in neurons of the mouse visual cortex. Aimed at  understanding and mimicking these results, we construct an artificial neural network and train it to classify images. Remarkably, we find that the best performance in such a task is obtained when the network operates near the critical point, at which the eigenspectrum of the covariance matrix follows the very same statistics as actual neurons do. Thus, we conclude that operating near criticality can also have ---besides the usually alleged virtues--- the advantage of allowing for flexible, robust and efficient input representations.
\end{abstract}


\section{Introduction}
Understanding how the brain of mammals, including humans, represents, processes and stores information is one of the main challenges of contemporary science. In addition to the obvious direct interest in such an ambitious goal, any progress made towards elucidating the brain working principles would also help developing a new generation of artificial intelligence devises. Reversely, advances in computer science help shedding light on the analogies and differences between our present operational  knowledge on "artificial intelligence· and  "natural intelligence". This two-sided dialogue is hoped to guide exciting breakthroughs in the next coming years in both fields. 

A popular idea, coming from the world of artificial neural networks \citep{Langton,Mitchell} and then exported to the realm of biological systems (see \citep{RMP} and refs. therein), is that information-processing complex systems, composed of many individual interacting units, are best suited to encode, respond, process, and store information  if they operate in a dynamical regime nearby the critical point of a phase transition, i.e. at the edge between "order" and "disorder"\citep{Mora-Bialek,Plenz-Functional,RMP,Chialvo2010,KC,Shriki,Breakspear-review,Shew2015b,Serena-LG,Martinello}.  In a nutshell, one can say that "ordered phases" encode information in  a robust or stable way, but they are not flexible enough as to accommodate for or respond to input changes; on the other hand, "disordered phases" are dominated by noise, thus hindering information storage and making retrieval exceedingly difficult.  Therefore, there needs to be some kind of  trade-off between order and disorder that can be formulated in a number of different ways, e.g., between "stability and responsiveness" or between "robustness and flexibility". 
The criticality hypothesis poses that such a trade-off is best resolved near criticality or "at the edge of chaos", where combined advantages from the two alternative phases can be obtained \citep{RMP}. Furthermore, at  critical points there is a concomitant scale invariance  ---with its characteristic power-law distributions and scaling---  entailing the existence of broadly different time and length scales, which seem much convenient for the representation of multiscale complex inputs.  Let us remark, that the terms "criticality" and "edge of chaos" are sometimes used indistinctly, though the last one applies to deterministic systems, in which a transition occurs between ordered and chaotic states, but as recently emphasized, they can describe two  sides of the same coin (we refer to \citep{Moritz} for an illuminating recent discussion).  

Empirical evidence that actual neural networks might operate close to criticality has kept accumulating in recent years \citep{BP,Petermann,Taglia,Plenz-synchro1,Shew2015b}. Most of this evidence (though not all) relies on the concept of neuronal avalanches \citep{BP} which are empirically observed to be scale invariant across species, brain regions, resolution levels and experimental techniques \citep{Schuster,Breakspear-review,RMP}. However, as of today, smoking-gun evidence is still needed to validate or dismiss this fascinating conjecture, and it remains controversial \citep{Touboul}; thus, novel theoretical and data-oriented analyses are much needed \citep{RMP}.


In a seemimgly-unrelated remarkable work, Stringer {\emph et al.} have recently made a step forward in understanding how neuronal networks actually represent complex inputs. In particular, these authors proved mathematically that the statistics of spiking neurons representing external sensory inputs (such as natural images represented in the mouse visual cortex) need to obey certain constraints for the input representation (or "neural code") to be "continuous and differentiable" \citep{Stringer}. These abstract mathematical properties are the formal counterpart of a much-desired property of neural networks: i.e., robustness of the representation against small perturbations of the inputs. Such 
a robustness is well-known to be often violated in artificial neural networks (ANNs); in particular, so-called, \emph{adversarial attacks}, consisting in  tiny variations in the input or their statistics, can fool the network, leading to wrong predictions and mis-classifications \citep{adversarial}.
We refer to Stringer \emph{et al.} \citep{Stringer} for an in-depth explanation and justification of these important ideas,  as well as to \citep{nassar_1n_2020} for an application of them onto multi-layer ANNs. In any case, the conclusion of Stringer \emph{et al.} is that, in order to achieve robust input representations, the covariance matrix of neuronal  activities measured across time when the network is exposed  to a sequential series of inputs, must obey the following spectral property: its rank-ordered eigenvalues should decay as a power law of their rank, with an exponent $\alpha$ strictly larger than $1+2/d$, where $d$ is the embedding dimension of the input. Thus,  $\alpha=1$ sets a lower bound for the possible values of the eigenspectrum decay-exponent for complex, high-dimensional inputs.

Rather remarkably, these theoretical predictions are verified to be fulfilled in experimental  recordings of more than $10000$ individual neurons in the mouse visual cortex exposed to a very large sequence  of natural images. This confirms that information encoding occurs in as mathematically predicted, i.e. in a continuous and differentiable manifold.
 
The main question we pose here is: are the internal representations of ANNs trained to classify images  similar to those of the mouse visual cortex? More specifically, is the spectrum of eigenvalues of the associated covariance matrix a power law of the rank? Is the exponent in all cases larger than (and close to) $1$? If so, do the exponent values change with the images dimensionality in the way predicted by Stringer et al.? 

Here, in order to tackle all these questions within the simplest possible scenario, we analyze the neural encoding of inputs with different dimensions in a paradigmatic example of ANN: the \emph{Echo state network} (ESN)  \citep{jaeger__2001}. This type of networks, together with \emph{liquid state machines} \citep{Maass,maass_real-time_2002}, constitute the prototype of \emph{reservoir computing} (RC) approaches \citep{lukosevicius_reservoir_2009}, a paradigm of computation that seems particularly well suited for exploiting the putative advantages of operating at the "edge of chaos" \citep{RMP}.

\section{Materials and Methods}
 The Echo State Network, in its original formulation, was devised by Jaeger as a flexible and easy-trainable recurrent neural network (RNN) for time-series prediction tasks \citep{jaeger__2001, jaeger_harnessing_2004}. More specifically, the architecture of ESNs is described as consisting of: 

\begin{itemize}

\item An input layer, which scales a number $L_{1}$ of inputs at each time step before they arrive in the reservoir, according to some random weights  $W^{in}\in\mathbb{R}^{N\times L_{1}}$.

\item A reservoir consisting of $N$ internal units connected with random weights  $W^{res}\in\mathbb{R}^{N\times N}$, whose states usually evolve according to a non-linear, time-discrete dynamical equation under the influence of a time-dependent input. In this way, the reservoir maps the external input into a high-dimensional space.  

\item An output layer, that converts the information contained in the high-dimensional states of the neurons (which serve as an internal representation of the inputs) to generate the final output.
\end{itemize}
Thus, unlike in other ANNs, the internal weights or "synaptic connections" in ESNs do not need to be updated during the learning process, and training is achieved by just modifying the layer of output weights that readout the network internal states. 

\begin{figure*}
\begin{centering}
\includegraphics[scale=0.5]{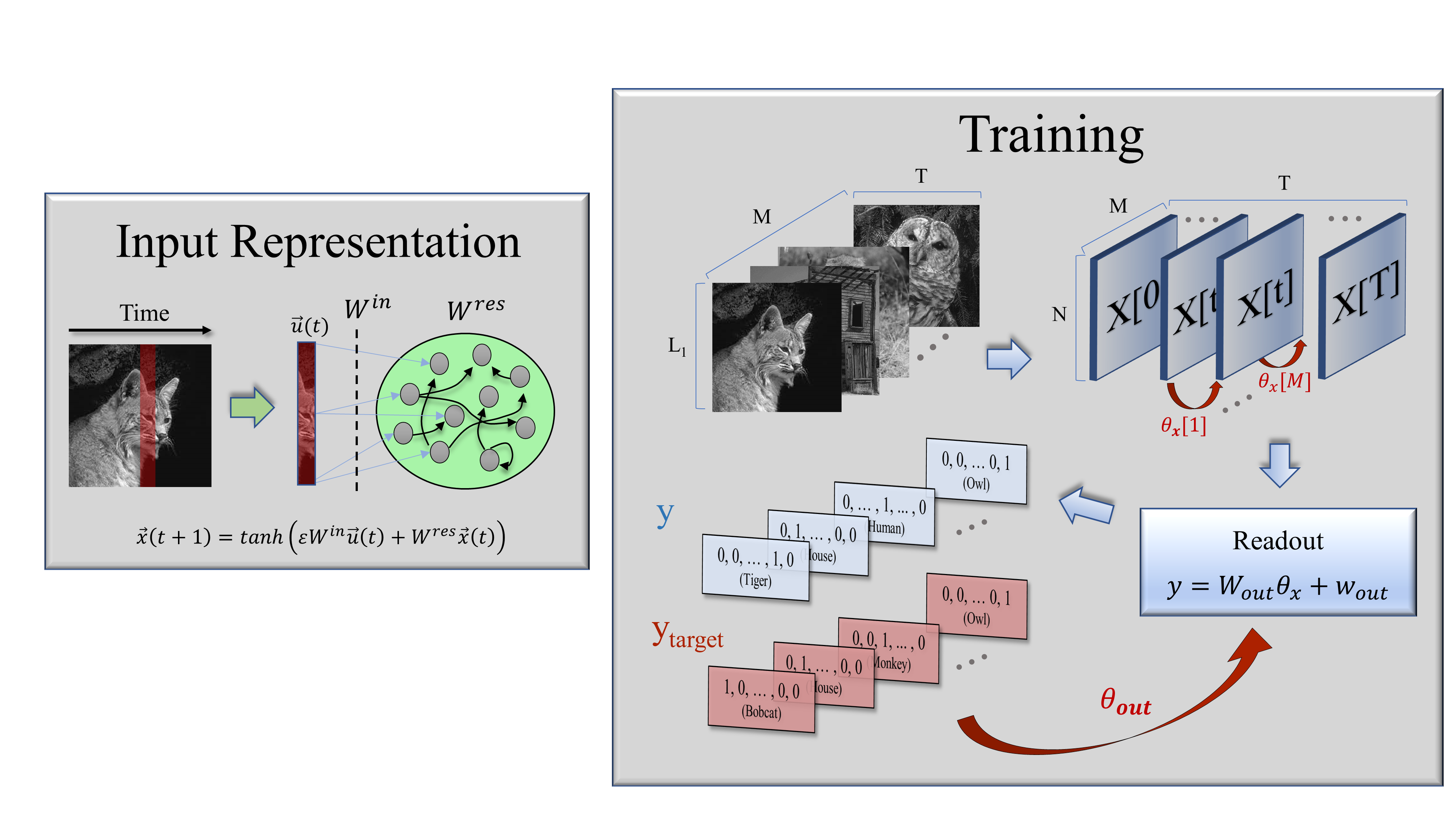}
\par\end{centering}
\caption{{\bf Sketch of the Echo State Network and the image classification task.} Left: Images are converted to multivariate time series and then fed into the reservoir. Right: for each processed image a set of parameters $\theta_{x}$ is generated, which characterizes the high-dimensional state of the reservoir, i.e. the ``\emph{reservoir model space}''. These are then fed into the readout module, that linearly transforms the information in the reservoir model space into an output label. Finally, output weights $\tilde{W}_{out}$ are generated by minimizing the error between the predicted and target labels. Red arrows indicate steps in which a Ridge regression is performed.}\label{Figure_0}
\end{figure*}

In order to adapt this architecture ---usually employed in time-series analyses--- for image classification tasks, we used black and white images with $L_{1}\times L_{2}$ pixels (each of them characterized by a value in the $[0,1]$ interval, representing a normalized gray-scale) and  converted them into multivariate time series by considering their vertical dimension as a vector of  $L_{1}$ elements or features, that "evolve" along $T=L_{2}$ discrete "time" steps. One can then define a standard training protocol in which, as illustrated in Fig.\ref{Figure_0}, at each time $t \in[0,T]$, vectors $\mathbf{u}(t) \in[0,1]^{L_{1}}$ corresponding to columns of the image are fed as inputs to the ESN. In this way, the network dynamics for the reservoir states is  given by the following non-linear activation function: 
\begin{equation}
\mathbf{x}(t)=\tanh(\varepsilon W^{in}\mathbf{u}(t)+W^{res}\mathbf{x}(t-1))\label{eq:States_Update}
\end{equation}
where $\varepsilon$ is an input scaling factor. 
Using a supervised learning scheme, the goal of the ESN is to generate an output label $\mathbf{y}\in\mathbb{\mathbb{N}}^{F}$ that correctly classifies each image in the test set as belonging to one of the $F$ existing categories or classes. This label consists of a vector in which every element is zero, except for a value of one at the position corresponding to the assigned class (i.e., "one-hot-encoded" in the machine learning jargon). Several readout methods have been proposed in the literature to transform the information contained in the reservoir dynamics into the expected target output $\mathbf{y}^{target}\in\mathbb{\mathbb{N}}^{F}$, ranging from linear regressions methods over the reservoir states \citep{reinhart_constrained_2010,reinhart_reservoir_2011}, to the use of "support vector machines" or "multilayer perceptrons as decoders \citep{babinec_merging_2006}. Here, we use a simple Ridge regression (see Appendix I for a detailed explanation of the algorithm) over the \emph{"reservoir model space"}, a method that has been recently proposed for the classification of multivariate-time series \citep{bianchi_reservoir_2021}. 

The "reservoir model space" is a set of parameters $\theta_{x}$ that encodes the reservoir \emph{dynamical} state for a given input (image). Such parameters are obtained from a Ridge regression to predict the next reservoir state from the past one at discrete time steps,
\begin{equation}
\mathbf{x}(t+1)=W_{x}\mathbf{x}(t)+\mathbf{w}_{x},
\end{equation}
in such a way that $\theta_{x}=\left[\mathrm{vec}(W_{x});\mathbf{w}_{x}\right]\in\mathbb{R}^{N\left(N+1\right)}$  provides a characterization of  the internal reservoir dynamical state during the presentation of a given input, where $\mathrm{vec}(\cdot)$ denotes reshaping to a one-column vector and $";"$  vertical concatenation. Then, for each image a readout module or decoder can transform this internal representation into an output label:
\begin{equation}
\mathbf{y}=W_{out}\theta_{x}+\mathbf{w}_{out}.\label{Eq_Readout}
\end{equation}
The parameters $\theta_{out}=\left[\mathrm{vec}\left(W_{out}\right);\mathbf{w}_{out}\right]$ ---where $W_{out}\in\mathbb{\mathbb{R}}^{F\times N(N+1)}$ and $\mathbf{w}_{out}\in\mathbb{\mathbb{R}}^{F}$ are defined as output weights and biases, respectively--- are determined again through Ridge regression, minimizing the error between the produced and target label for all the presented images in the training set. 

Let us remark that the presented framework can be naturally extended to include, for instance, leakage and noise terms in Eq.\ref{eq:States_Update}, feedback connections from the output to the reservoir, or plastic rules that modify the reservoir weights according to the inputs  \citep{lukosevicius_practical_2012,morales_unveiling_2021}, among other possible extensions. However and since our aim here is not to reach state-of-the-art classification accuracy ---but rather highlight the link between optimal input representation and the internal dynamical state--- for the sake of parsimony, we refrain from adding further features to our model for the time being.

\section{Results}

\subsection{Non-trivial scaling at the edge of chaos.}
Although relatively simple, our proposed ESN model has several hyperparameters than can be tuned, affecting its performance. More specifically, the \emph{spectral radius} $\rho$ of the reservoir internal weight matrix and the \emph{scaling factor} $\varepsilon$ of the input weights are two variables that usually determine the dynamical regime within the reservoir \citep{lukosevicius_reservoir_2009}. The spectral radius ---or largest eigenvalue of the reservoir weight matrix--- controls the dynamical stability inside the reservoir when no input is fed into the network. Thus, a spectral radius exceeding unity has been often regarded as a source of instability in ESNs due to the loss of the so-called \textquotedblleft \emph{echo state property}\textquotedblright, a mathematical condition ensuring that the effect of initial conditions on the reservoir states fades away asymptotically in time \citep{jaeger__2001,jaeger_short_2001,yildiz_re-visiting_2012}. Nevertheless, later studies have shown that the echo state property can be actually maintained over a unitary spectral radius, and different sufficient conditions have been proposed  \citep{buehner_tighter_2006,yildiz_re-visiting_2012,gallicchio_chasing_2018} (see in particular \citep{manjunath_echo_2013}, where the authors analyze the problem from the lens of non-autonomous dynamical systems, deriving a sufficient condition for the echo state property with regard to a given input). On the other hand, increasing the value of  $\varepsilon$  can convert an initially expanding mapping into a contracting dynamics, as stronger inputs tend to push the activities of the reservoir units towards the tails of the non-linearity.

In what follows,  we analyze the input representation that the reservoir codifies in terms of the trade-off between $\rho$ and $\varepsilon$, which \emph{together} determine the dynamical operating regime of the ESN and the presence or absence of the echo state property. For the rest of parameters, the number of units in the reservoir is kept fixed to $N=2000$ and the density of the reservoir-weight-matrix elements (i.e., the percentage of non-zero connections) to $10\%$, while both reservoir and input weights are extracted at random from a uniform distribution in the interval  $[-1,1]$. 
\begin{figure}
\centering{}\includegraphics[scale=0.5]{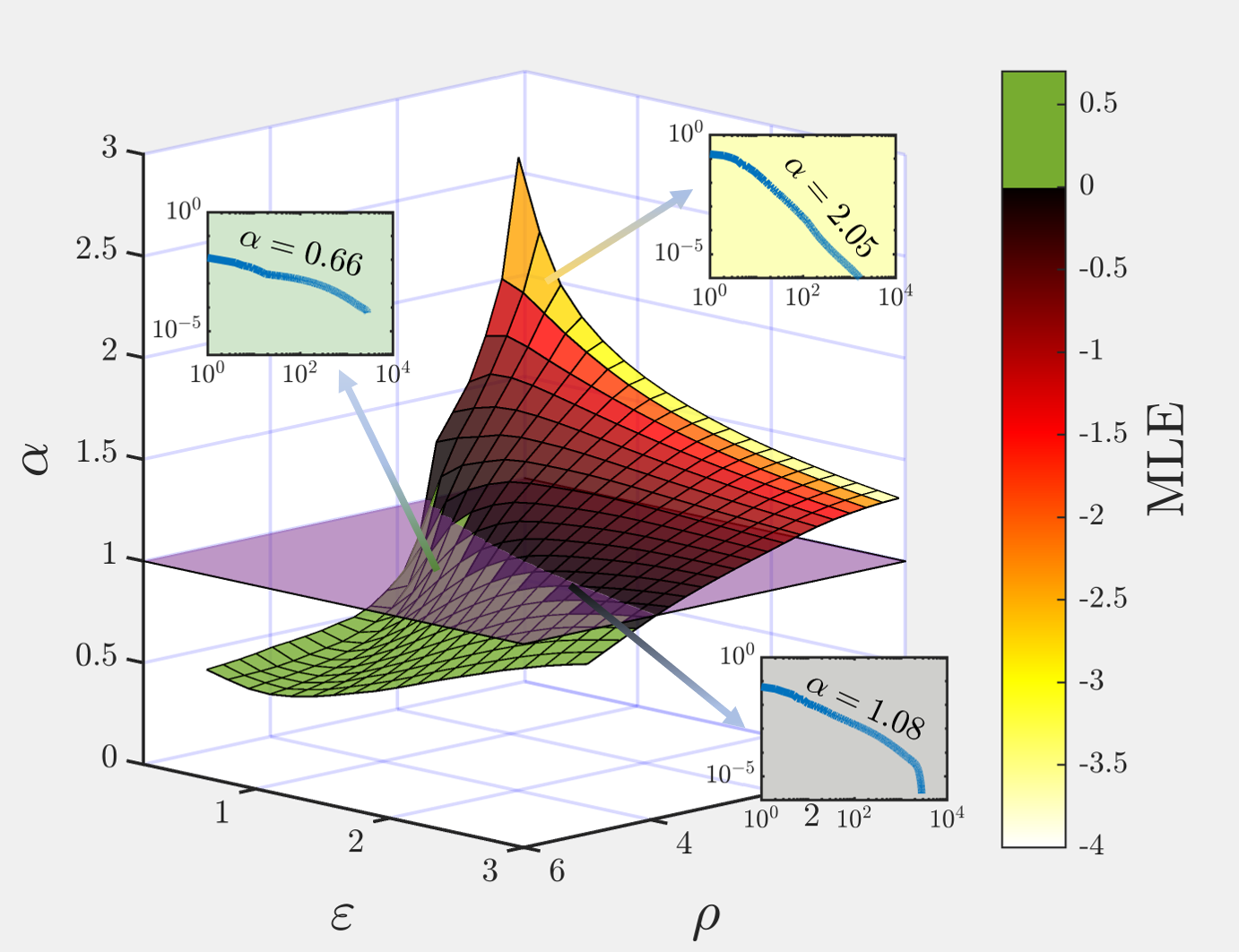}\caption{Exponent for the power-law decay of the spectrum of the activity covariance matrix  as a function of the spectral radius ($\rho$) and input scaling factor ($\varepsilon$) of the reservoir, plotted together with the maximum Lyapunov exponent (MLE) color-coded within the surface. The insets correspond to the activity covariance matrix eigenspectrum measured in three different points of the parameter space, where the variance in the n-th dimension (n-th eigenvalue) scales as a power-law  $n^{-\alpha}$  of the rank.  For ease of visualization, the plane separating the region $\alpha < 1$  in which the representation is no longer continuous nor differentiable was plotted in purple.}\label{Figure_1}
\end{figure}

Following the same methodology as in Stringer {\emph et al.}\citep{Stringer}, the ESN was first presented with a large set of high-dimensional, natural images, and the activity of the internal units in the reservoir  was stored for each step of the training. Then, principal component analysis (PCA) was performed directly over the full set of neuron activities $X\in\mathbb{R}^{N\times( T\times M)}$,  where $T=90$ is the number of pixels in the horizontal dimension of the images and
$M=2800$  is the total number of images.  In this way, we obtained the variance along each principal component or eigenvector of the covariance matrix, that serves as a basis for the activity inside the reservoir (see \citep{shlens_tutorial_2014} for a very gentle but rigorous introduction to PCA).

Notably, we found that the spectrum of eigenvalues as a function of their rank (i.e., the variance associated to the n-th principal component, when ordered from the largest to the smallest) can be well fitted to a power-law  $n^{-\alpha} $ (see insets in Fig.\ref{Figure_1}), whose  associated exponent $\alpha$ decays (flatter spectrum) with the spectral radius $\rho$; 
while, the exponent increases (faster decay) with the input scaling factor $\varepsilon$  for most of the parameter space (see Fig. 2). In \citep{Stringer} it was found that the exponent of this power-law relation is close to $1$  when natural, high-dimensional images were shown to the mouse as an input. As discussed above, the authors proved mathematically that $\alpha>1+2/d$ is a necessary condition for the neural manifold that emerges from the representation of a $d$-dimensional input to be continuous and differentiable. For natural images, $d$  is very large and one can approximate the critical exponent by $\alpha_{c}\approx1$ (this condition is marked by the purple plane in Fig.\ref{Figure_1}). 

As a following step, one might  naturally wonder if there is some
 aspect of our model that characterizes such a regime of robust representations in the parameter space $(\rho,\varepsilon)$, for which an exponent $\alpha$ close to unity is found. In other words: is the dynamics of the system inherently different in the regions for which the input representation manifold is found to be non-analytical (i.e. below the purple plane in Fig.\ref{Figure_1})?

As it turns out, rather remarkably, the exponent $\alpha$ characterizing the decay of the eigenspectrum approaches unity for choices of $\rho$ and $\varepsilon$ that drive the network dynamics towards the so-called "edge-
 of instability" or "edge of chaos", this is, near a transition point between an ordered and a chaotic regime. Traditionally, chaotic regimes are  characterized by their average sensitivity to perturbations in the initial conditions; to quantify this effect, one usually measures the rate of divergence of two trajectories with a very small difference in their initial conditions:
\begin{equation}
\lambda=\lim_{k\to\infty}\dfrac{1}{k}log\left(\dfrac{\gamma_{k}}{\gamma_{0}}\right)
\end{equation}where $\lambda$ is termed the maximum Lyapunov exponent (MLE), $\gamma_{0}$ is the initial distance between the perturbed and unperturbed trajectories, and $\gamma_{k}$ is the distance between the trajectories at step $k$  (we refer the reader to \citep{boedecker_information_2011} and \citep{sprott_chaos_2003} for a detailed explanation of the algorithm used to compute the MLE). Thus, chaotic dynamics is typically associated with a positive MLE, while the system is said to be stable to local perturbations provided $\lambda<0$. It can be clearly seen from Fig.\ref{Figure_1} that the region in which one finds non-analytical representations of the input (below the purple plane) matches almost perfectly with the region (colored in green) in which a positive MLE is found. 

The transition order-to-chaos can be also visualized looking directly at the activities inside the reservoir (see Fig.\ref{Figure_5}). Observe that, when the network is in an "ordered" state, with $\lambda<0$, the responses of the neurons are quite heterogeneous when compared among them, but they are highly localized within each neuron, i.e., individual neurons experience a limited response to stimuli. On the other hand, dynamical states characterized by $\lambda>0$ have neurons whose response extends across the full range of the non-linearity (with higher probability along the tails, reflecting a saturated behavior), but it is this same "phase space expansion" that makes units almost indistinguishable from each other. It is only around the critical point or f-chaos, that we find a trade-off between dynamical richness in individual units and variability across units. 

\begin{figure}
\centering{}\includegraphics[scale=0.8]{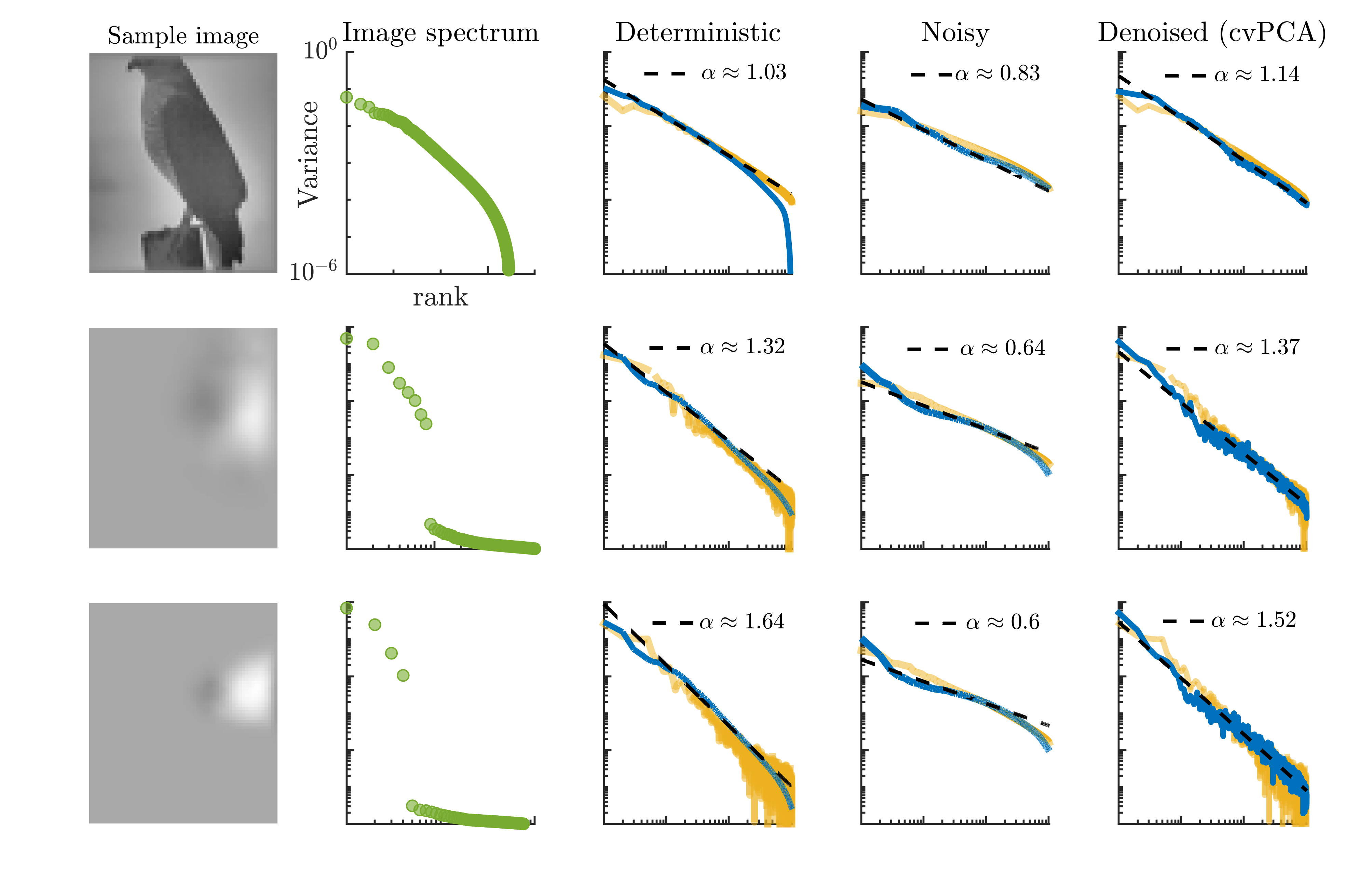}\caption{From left to right: (A) sample from the $M=2800$  images in the training set; (B) eigenspectrum of the images pixel intensities; (C) eigenspectrum for the activities of an ESN (blue line) and real, V1 mouse neurons (yellow line, plotted after  \citep{Stringer} applying cvPCA) when subject to images of dimensionality $d$ ; (D) same analysis, but now zero-centered white noise of amplitude $\epsilon = 0.4$  is added to the neuron dynamics (blue line), and no cvPCA is performed over the experimental values (yellow line); (E) same analysis as in (D), but now noise has been substracted using cvPCA. From top to bottom: results for natural, high-dimensional images; the same images projected onto 8 dimensions; the same images projected onto 4 dimensions. To obtain the ESNs eigenspectra, parameters were chosen so that the networks operated near the edge of chaos, with $\lambda \sim -5 \cdot 10^{-3}$. }\label{Figure_2}
\end{figure}

Coming back to the results in Stringer \emph{et al.}, one may also wonder whether the continuity and differentiability condition $\alpha>1+2/d$ holds also for low-dimensional inputs, for which the expected bound $\alpha_{c}=1+2/d$  deviates considerably from unity.  To this purpose, Fig.\ref{Figure_2} shows the measured eigenspectrum of the reservoir activity covariance matrix (i.e. eigenvalues as a function of their rank) when images of different dimensionality (the same ones used by Stringer \emph{et al.} in their experiments) are presented as inputs, and the reservoir is tuned to operate at the onset of a chaotic regime, i.e., for values in the parameter space $\left(\rho,\varepsilon\right)$ for which $\lambda$ was near zero but still negative. Remarkably, we find in all cases that \emph{the exponents observed in the mouse visual-cortex activity are best reproduced when the reservoir dynamics is tuned close to the "edge of chaos"}. 

This finding suggests that one can set the network parameters in such a way that the neural activity manifold in which the input is represented is almost as high-dimensional as possible without loosing its "smoothness", and that such optimal solution is found at the edge of chaos.

At this point, it is pertinent and timely to dig a bit deeper on the similarities and differences between the results presented in \citep{Stringer} for real, V1-cortex neurons in the mouse, and the power-law exponents obtained through our reservoir computing model. 

(i) First of all, as in the case of real neurons, the observed correlations between the internal units are  not just a byproduct emerging from scale-free features of natural images (see second column in Fig.\ref{Figure_2}). In particular,  one can see that the power-law decay of the eigenspectrum persists even in response to low-dimensional inputs whose embedding vector space can be spanned with just a few principal components (i.e. without a power-law decaying intrinsic spectrum).

(ii) In our model, images are processed sequentially in time along their horizontal dimension, so that for each image one can measure the activity of the $N$ internal units over $T=L_{2}$ time steps. In contrast, activity of V1 neurons in \citep{Stringer} is scanned at a relatively low rate, so that for each image the neural representation in characterized by just one amplitude value in each neuron. 

(iii) To avoid confusion, let us remark that the variance observed by Stringer et al. is not directly measured over the raw activity of the neurons. Instead, the authors first project out the network spontaneous activity from the data, and then perform a cross-validated PCA (cvPCA) that allows them to filter out the trial-to-trial variability or "noise". The cvPCA method is thus able to estimate the stimulus-related variance confined in an n-dimensional manifold by first computing the eigenvectors spanning this manifold from a first repeat of the full training set, and then measuring the amount of a second repeat's variance that is confined to this plane (we refer to \citep{Stringer} for a detailed explanation and derivation of the cvPCA method). However, since our model is completely deterministic for a given initialization of an ESN, the stimulus-related variance computed through cvPCA  trivially matches that of a standard PCA.

A natural question then arises from this last point: what happens when a noise term is included in Eq.\ref{eq:States_Update}, so that the dynamics becomes stochastic? Are the  power-law exponents robust to the introduction of noise? To answer these questions, we considered stochastic versions of the ESNs  ---including an independent small additive noise term in their inputs--- and presented them with two repeats of the same input training set. The internal states of the noisy  reservoirs were collected at each time step. We then performed the same type of cvPCA analyses proposed in \citep{Stringer} to estimate the signal variance in our reservoirs (see fourth and fifth columns in Fig.\ref{Figure_2}). Just as in the case  of real V1 neurons, the exponents measured over the raw, noisy activity are lower and below the critical threshold for continuity and differentiability of the neural manifold. Nevertheless, a cvPCA over the internal states retrieves the expected exponents after noise has been filtered out.
We will further comment on the possible implications of this finding in the Discussion section, but for now, let us wrap up our findings tackling what we believe is a fundamental question from the perspective of machine learning: does working at the edge of chaos (or, equivalently, having maximal, continuous and differentiable neural manifolds) provide any functional advantage?

\subsection{Solving a bench-mark classification task.}

The advantages of working at the so-called edge of chaos were first pointed in general dynamical systems and cellular automata \citep{crutchfield_computation_1988,Langton}, and only later analyzed in reservoir computer models with binary \citep{bertschinger_real-time_2004, legenstein_edge_2007, busing_connectivity_2010} and analog \citep{busing_connectivity_2010, boedecker_information_2011} internal units. In particular, in \citep{boedecker_information_2011} the authors showed that ESNs presented maximal information storage and transfer, as well as enhanced memory capacity right at the edge of chaos. However, while ESNs and other RC approaches have been previously applied to classification tasks with very good results \citep{schaetti_echo_2016, skowronski_automatic_2007, aswolinskiy_time_2016, ma_functional_2016, yusoff_modeling_2016,bianchi_reservoir_2021}, to the best of our knowledge an analysis of the influence of the dynamical regime on the performance of RC architectures for classification tasks is still missing. To this end, we measure the performance of ESNs in a classification task over the canonical MNIST dataset, which includes $60,000$ handwritten instances of  the first $10$ digits as training (although only a third of them was used) and $10,000$  for testing, following the training and testing procedure as described in Materials and methods. 

\begin{figure}
\centering{}\includegraphics[scale=1]{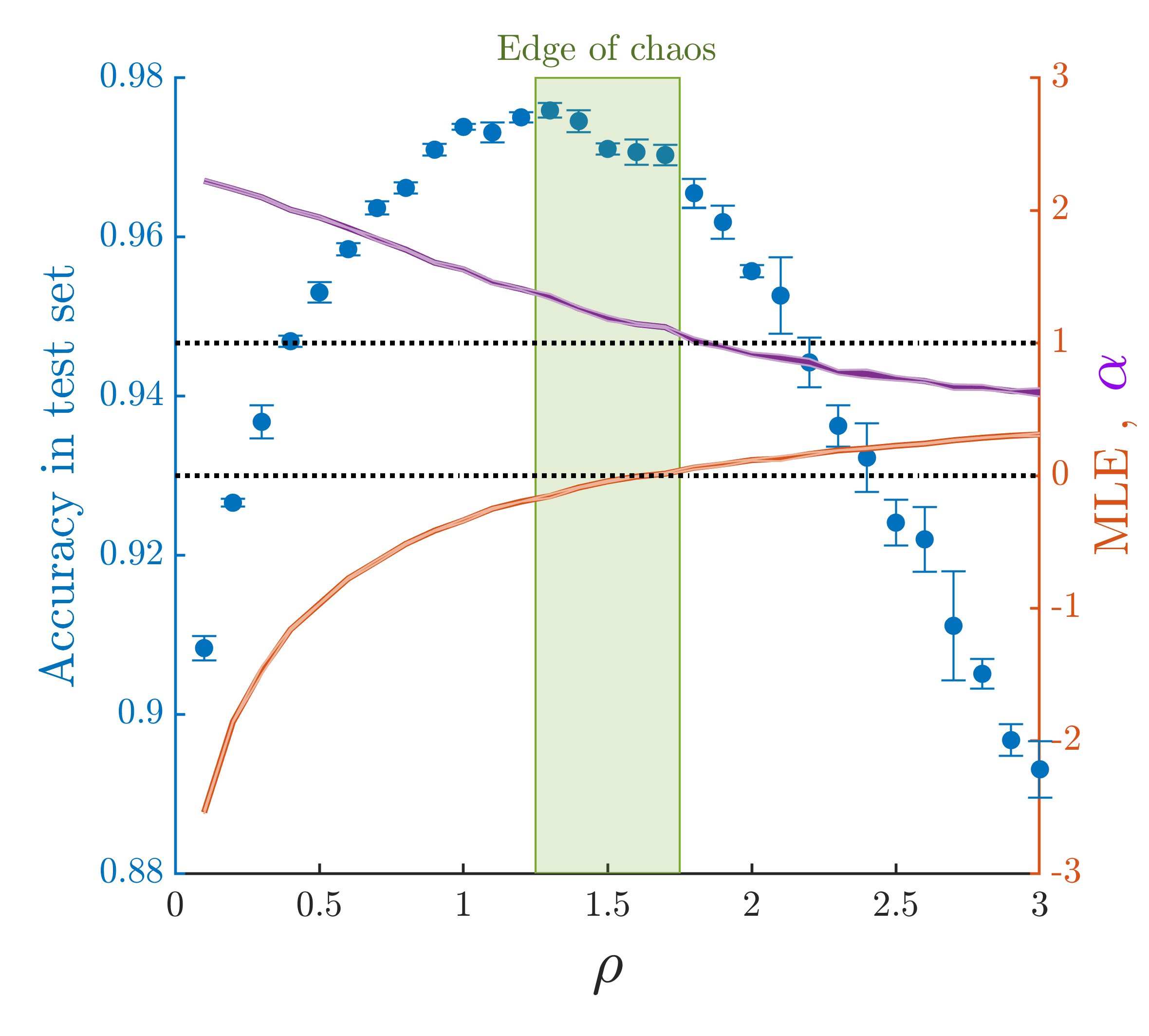}\caption{Curves for the accuracy in MNIST testset (blue dots), maximum Lyapunov
exponent (orange line) and best fit exponent for power-law spectrum
of activity covariance matrix (purple line). Training was performed
over 20,000 randomly chosen images of the MNIST training dataset, while classification error was assessed
over the full test set (10,000 images). Errors in each case were estimated as the standard deviation from the mean over ten different initializations of the ESN.}\label{Figure_4}
\end{figure}

The results, showed in Fig.\ref{Figure_4}, highlight the fact that optimal performance ($\sim 2.2 \%$ error rate) is found just below the onset of chaos, when $\lambda \lesssim 0$. Most notably, the plot also evinces that the decay in performance is not only preceded by a positive MLE, but coincides too with exponents $\alpha$  for the fit of the covariance-matrix eigenpectrum that are below the  limiting value $\alpha_{c}\approx1$, indicating the loss of continuity and differentiability of the neural representation manifold for high-dimensional images. Let us remark that, the slower the decay (i.e. the larger the exponent) the more weight is given to fine details of the input, but if the decay is too slow (smaller than the lower bound above), an excessive importance is given to such fine details at the cost of hampering the existence of a "smooth manifold" representation. Thus, operating near the edge of instability could provide the network with an optimal trade-off between representing as much details as possible and constructing operative, smooth representations.

We finally remark that the results shown here were obtained with a reservoir consisting only of $500$ internal units and using only one-third of the training set, with no pre-processing of the images. In contrast, the current best performance in MNIST digit recognition ($0.81\%$ error rate) using reservoir computing networks has been achieved with a two-layer architecture, each consisting of $16,000$ units, which amounts to a total $880,000$ trainable parameters \citep{jalalvand_design_2015}. In this sense, a simple ESN with readouts over the reservoir model space, when tuned near the edge of chaos, is able to outperform ESNs with a greater number of units and much more complicated dynamics (including feedback connections and leakage terms), trained over the full MNIST dataset \citep{schaetti_echo_2016}. 

\section{Discussion}
 Stringer {\emph{et al.}} observed in  \citep{Stringer} that neural coding of different inputs in the mouse V1-cortex is close to optimal for each type of input, constrained by requirements of continuity and differentiability of the neural response manifold. In this paper, we open the door to the possibility that optimal, continuous and differentiable response manifolds emerge for neuron dynamics laying close to an edge of chaos type of critical point. Indeed, we have shown that a simple non-linear model of  randomly-connected neurons, when subject to an external input, is able to reproduce power-law exponents similar to those found in mouse V1-cortex for the decay of the covariance matrix eigenspectrum.

We find nevertheless important to clarify that the term edge of chaos ---and the concept of chaos itself--- should be taken with caution as it is not devoid of criticism in this context.  As pointed out in \citep{manjunath_echo_2013}, ESNs are an example of nonautonomous dynamical systems, for which typical concepts based in the theory of autonomous systems (e.g., "sensitivity to initial conditions", "attractor" and "deterministic chaos") do not directly apply \citep{clemson_discerning_2014,gandhi_theory_2012}. In fact, the authors  of \citep{manjunath_echo_2013} claim that local perturbation experiments cannot represent an ultimate evidence of chaotic dynamics  in non-autonomous systems, since it might well be the case that the input drives the system towards and expanding dynamics for a certain time span, while the system shows on average a contracting, non-chaotic dynamics. Despite these caveats, at the light of the presented results it appears like there is indeed an actual dynamical phase transition occurring as the maximum Lyapunov exponent crosses zero. Thus, in any case, it seems a sensible choice to use such a quantity as a control parameter when analyzing the underlying neural representation of external inputs. 

On the other hand, adding stochasticity in the form of small-amplitude white noise naturally leads to flatter eigenspectra, much like those found when PCA is performed over raw experimental data. Nevertheless, one can use the same cvPCA technique introduced also in \citep{Stringer} to extract the input-related variance of the activity, thus obtaining similar exponents to the fully deterministic case. This remarkable result therefore suggests that the role of spontaneous activity and trial-to-trial variability on the representation of external inputs can be easily accounted for in our simple echo-state-network model.

Finally, results obtained on a benchmark classification task suggest that input-representation manifolds that are critically high-dimensional (from the point of view of their analytic properties) may serve a bigger purpose than just being a mathematical curiosity, as ESNs show a better performance when poised near such a critical point, while the accuracy falls rapidly as soon as the representation manifold becomes fractal.

Therefore, the presented results open the path for very exciting research avenues at the boundary of biology and machine learning, calling for theoretical formulations that can shed light into the fascinating properties of these input-representation neural manifolds and their relation with the criticality hypothesis.

\section*{Appendix A: Ridge regression}

Readouts in ESNs are typically linear, as it can be seen from Eq.\ref{Eq_Readout}. To simplify the forthcoming derivations, let us rewrite Eq.\ref{Eq_Readout} as:
\begin{equation}
\mathbf{y}=\tilde{W}_{out}\left[\theta_{x};1\right]
\end{equation}
where $\tilde{W}_{out}\in\mathbb{\mathbb{R}}^{F\times\left(N(N+1)+1\right)}$ and " ;  " indicates vertical vector concatenation. Thus, for a given input image a reservoir representation of the states $\theta_{x}$ is constructed, and one can use the above equation to generate the corresponding output (which in our case will be a one-hot-encoded label $\mathbf{y}\in\mathbb{\mathbb{N}}^{F}$ classifying the image). We can trivially generalize the above equation to apply to the full set of $M$ images in the training set $Y\in\mathbb{\mathbb{N}}^{F\times M}$:
\begin{equation}
Y=\tilde{W}_{out}\Theta_{x}
\end{equation}
where $\Theta_{x}\in\mathbb{R}^{\left(N\left(N+1\right)+1\right)\times M}$ contains as columns the vectors $\left[\theta_{x};1\right]$ generated for each of the $M$ input images. Finding the optimal weights
$\tilde{W}_{out}$ that minimize the squared error between the produced and target labels, $\mathbf{y}$ and $\mathbf{y}^{target}$, is then reduced to a standard linear regression problem, which is the greatest strength of the reservoir-computing approach:
\begin{equation}
Y^{target}=\tilde{W}_{out}\Theta_{x}.
\end{equation}
Owing to the fact that large output weights are commonly associated to over-fitting of the training data \citep{lukosevicius_practical_2012}, it is a common practice to add a regularization term to the error in the target reconstruction, usually defined in terms of the root-mean-squared error (RMSE). Although several methods have been proposed to achieve this regularization \citep{lukosevicius_practical_2012, reinhart_constrained_2010,reinhart_reservoir_2011}, one of the most efficient and stable algorithms is Ridge regression, which aims to solve:
\begin{equation}
\tilde{W}_{out}=\underset{\left\{ \tilde{W}_{out}\right\} }{\arg\min}\dfrac{1}{M}\sum_{n=1}^{M}\sum_{i=1}^{F}\left(y_{i}[n]-y_{i}^{target}[n]\right)^{2}+\beta\left\Vert \tilde{w}_{i}^{out}\right\Vert ^{2}=Y^{target}\Theta_{x}^{T}\left(\Theta_{x}\Theta_{x}^{T}+\beta I\right)^{-1},\label{eq:MSE_Error}
\end{equation}
where $\left\Vert \text{\ensuremath{\cdot}}\right\Vert $ stands for the Euclidian norm, $I$ is the identity matrix and $\beta$ is the regularization coefficient. Notice that choosing $\beta=0$ removes the regularization, turning the Ridge regression into a standard generalized linear regression problem (we used $\beta = 1$ across all simulations in the paper).

We finally remark that, in order to obtain the reservoir model space parameters $\Theta_{x}$, the same Ridge regression is also implemented to solve Eq. 2 for each of the input images.

\section*{Appendix B: Phase space of reservoir units}
\begin{figure}
\centering{}\includegraphics[scale=1]{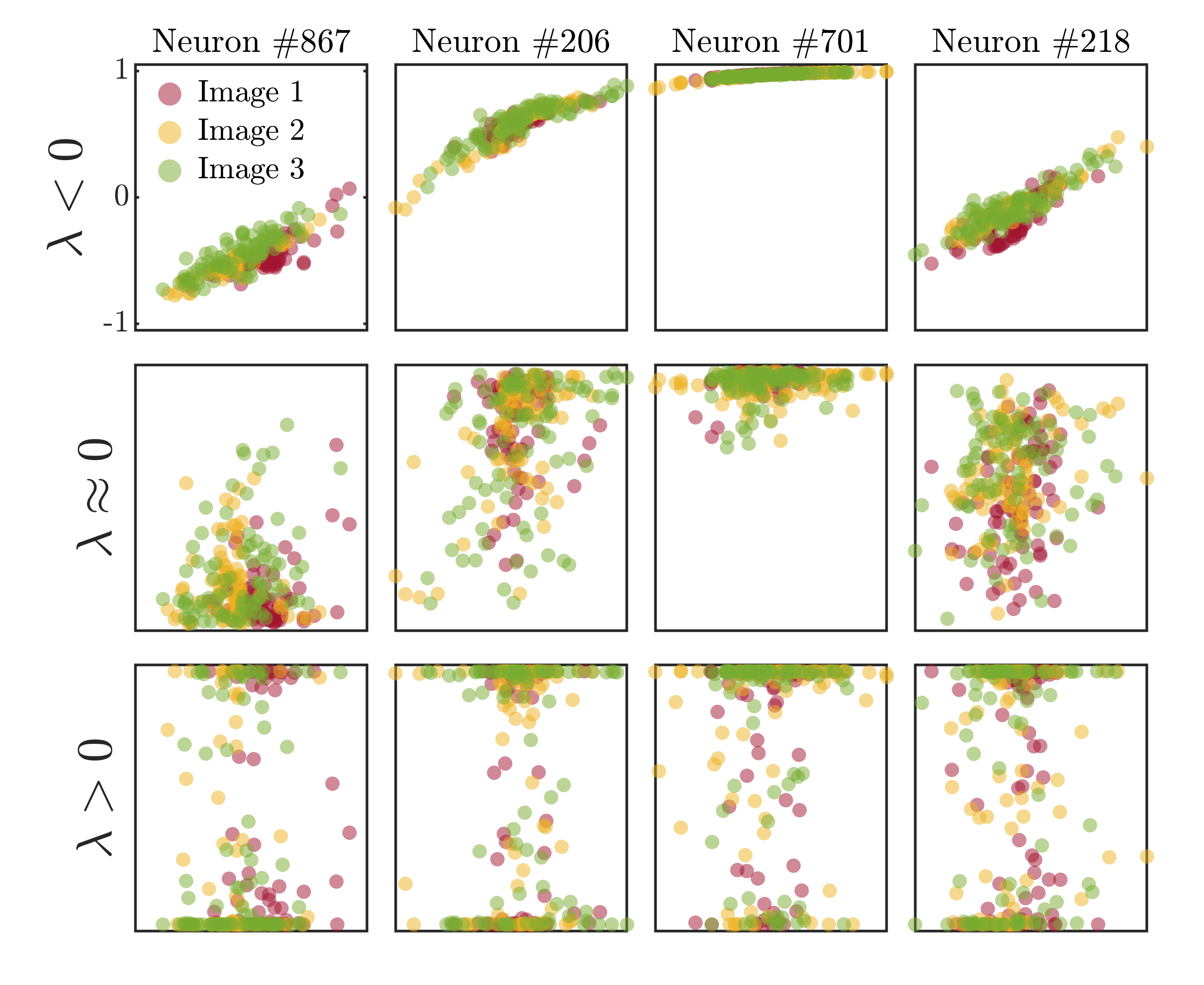}\caption{Activity of four different neurons operating in the sub-critical, critical and super-critical regimes (from top to bottom) when presented with three different high-dimensional, natural images. Each point in the panels represents the activity $x_{i}(t)$  of neuron $i$  at time step $t$  as a function of the total input $f_{i}(t)=\varepsilon \sum_{l=1}^{L_{1}} w_{il}^{in} u_{l}(t) + \sum_{j=1}^{N} w_{ij}^{res} x_{j}(t-1)$ arriving to it.} \label{Figure_5}
\end{figure}

In Fig.5, we reproduce the phase space of $4$ different internal units of a reservoir operating (from top to bottom lines) in a sub-critical, critical and super-critical regimes, respectively.  Each plot represents the activity of the corresponding neuron at each time step against the total input (sum of external input plus reverberating activity in Eq.\ref{eq:States_Update}) arriving to the neuron at the previous time. Since each image is first transformed into a multivariate time series of $T=L_{2}=90$ time steps ---and we plot the activity along the first three images only--- each panel in Fig.\ref{Figure_5} contains 270 points.

\bibliography{Reservoir_Computing_Bib}

\end{document}